\documentclass[draft]{agujournal2019}
\usepackage{url} 
\usepackage{lineno}
\usepackage[inline]{trackchanges} 
\usepackage{soul}
\usepackage{float}
\usepackage{amsmath}
\usepackage{amssymb}
\usepackage{comment}
\usepackage{xcolor}
\usepackage{caption}
\usepackage{subcaption}
\nolinenumbers
\draftfalse

\journalname{Enter journal name here}

\begin{document}

%
%


\title{Assessing the predicting power of GPS data for aftershocks forecasting}

%
%




\authors{Vincenzo Maria Schimmenti\affil{1,2}, Giuseppe Petrillo\affil{3}, Alberto Rosso\affil{1}, Francois P. Landes\affil{2}}

\affiliation{1}{Universit\'e Paris-Saclay, CNRS, LPTMS, 91405, Orsay, France}
\affiliation{2}{Universit\'e Paris-Saclay, CNRS, INRIA, Laboratoire Interdisciplinaire
des Sciences du Num\'erique, TAU team, 91190 Gif-sur-Yvette, France}
\affiliation{3}{The Institute of Statistical Mathematics, Research Organization of Information and Systems, Tokyo, Japan}





\correspondingauthor{Vincenzo Maria Schimmenti}{vincenzo.schimmenti@universite-paris-saclay.fr, vinci95s@gmail.com}




\begin{keypoints}
\item Building a dataset based on GPS data for aftershocks forecasting.
\item Using machine learning algorithms for aftershock forecasting.
\item Evaluating the usefulness of GPS data for seismic prediction purposes.
\end{keypoints}

%
%

%
%


\begin{abstract}
We present a machine learning approach for the aftershock forecasting of Japanese earthquake catalogue from 2015 to 2019. Our method takes as sole input the ground surface deformation as measured by Global Positioning System (GPS) stations at the day of the mainshock, and processes it with a Convolutional Neural Network (CNN), thus capturing the input's spatial correlations.
Despite the moderate amount of data the performance of this new approach is very promising. The accuracy of the prediction heavily relies on the density of GPS stations: the predictive power is lost when the mainshocks occur far from measurement stations, as in offshore regions.  
\end{abstract}

\section*{Plain Language Summary}
Forecasting large aftershocks is a challenge of great importance for human security. Today we dispose of statistical predictive models called Epidemic Type Aftershock Sequence (ETAS) tuned on the earthquake catalogue of the past seismicity. This catalogues contains basic information such as the location, the time and the magnitude of an earthquake. However we dispose of much richer data set about the crust dynamics, such as the daily displacement of the ground surface, that is nowadays measured by numerous GPS stations, devices that send their absolute position everyday to sattellites, thus telling us about how the ground deforms. In this study, we propose to forecast the Japanese aftershocks by means of a machine learning study of the GPS data alone. Our results show that this method is very promising and relies on the quality and the quantity of the available data.

%
%

%


%
%
%
%

\section{Introduction}
A large earthquake triggers numerous aftershocks with a frequency  that decreases with time according to the celebrated Omori law  \cite{utsu1995centenary}.
The aftershocks magnitude is typically small, but some of them can be very large and dangerous because they have occurred in areas that already weakened.
Forecasting their spatial location  remains an outstanding open problem. Today, for practical purposes this task is carried out using a statistical model called Epidemic-Type Aftershock Sequence (ETAS) model \cite{ogata1988,ogata1998,HS03,OZ98}. The ETAS is a self-exciting point process in which aftershocks occurrence is influenced by past events. Few free parameters define the model and they are calibrated by the simple catalog data containing magnitude, time and location of the events \cite{Lippiello2014,lombardi2015,molkenthin2022,ross2021}. This task is challenging due to the overlapping of seismic coda waves, which makes difficult to detect small events in the initial stages of seismic sequences, leading to a significant bias in the estimation of ETAS parameters \cite{kagan2004short,peng2007seismicity,omi2013forecasting,hainzl2016apparent,hainzl2016rate,de2018overlap,lippiello2019forecasting}. Recently, some fitting procedures have been developed to address this data incompleteness, but their effectiveness relies on the number of recorded aftershocks \cite{lippiello2019forecasting,omi2015intermediate,omi2019implementation}. Another limitation of these forecasting  methods is the need to convert the estimated aftershock rate into the observed ground shaking, via accurate empirical attenuation functions \cite{boore1982empirical,fukushima1990new}. 

A second strategy of aftershock forecasting relies on the study of  more rich and structured data such as the ground displacement map generated by the mainshock. This map is estimated from the inversion of the  waveform data by means of finite-fault rupture models designed by geophysicists for each seismic fault. The Coulomb fault stress change obtained from the map provides a simple and geophysically  inspired criterion for aftershock forecasting \cite{king1994static,toda1998stress,parsons1999stress,reasenberg1992response,jacques1996seismic, nostro1997static}. However its validity is still controversial \cite{hardebeck1998static,mallman2007assessing, felzer2005testing,petrillo2022testing} and cannot compete with the performance of ETAS based predictions (in particular ETAS systematically overperform on the standard CSEP test \cite{savran2022,csep,kh2023}). 
More recently machine learning approaches have been explored with more promising results. In this context neural networks, namely models defined by many free parameters, are trained using this more rich data set. Specifically the authors in \cite{DeVries2018} trained a Multilayer Perceptron (MLP) with the ground deformation maps to forecast aftershock locations of different large mainshocks around the world. As discussed above, these maps are not the outcome of a direct measurement but rely on the inversion with complex rupture kernels. In this work we explore the possibility to train a neural network using surface displacement maps recorded by GPS stations. We focus on the data of Japan (1998-2018) where the GPS stations are dense, with an average separation  of $11$ km, and a relatively good amount of independent sequences are available in the Japan Meteorological Agency Earthquake Catalog (JMA) \footnote{https://www.data.jma.go.jp/svd/eqev/data/bulletin}.

The paper is structured as follows: Section II provides an introduction to the data and the methods utilized. The results are presented in Section III, and the final section is dedicated to conclusions.

\section{Data and methods}

\begin{figure}[h]
    \centering
    \includegraphics[width=0.9\textwidth]{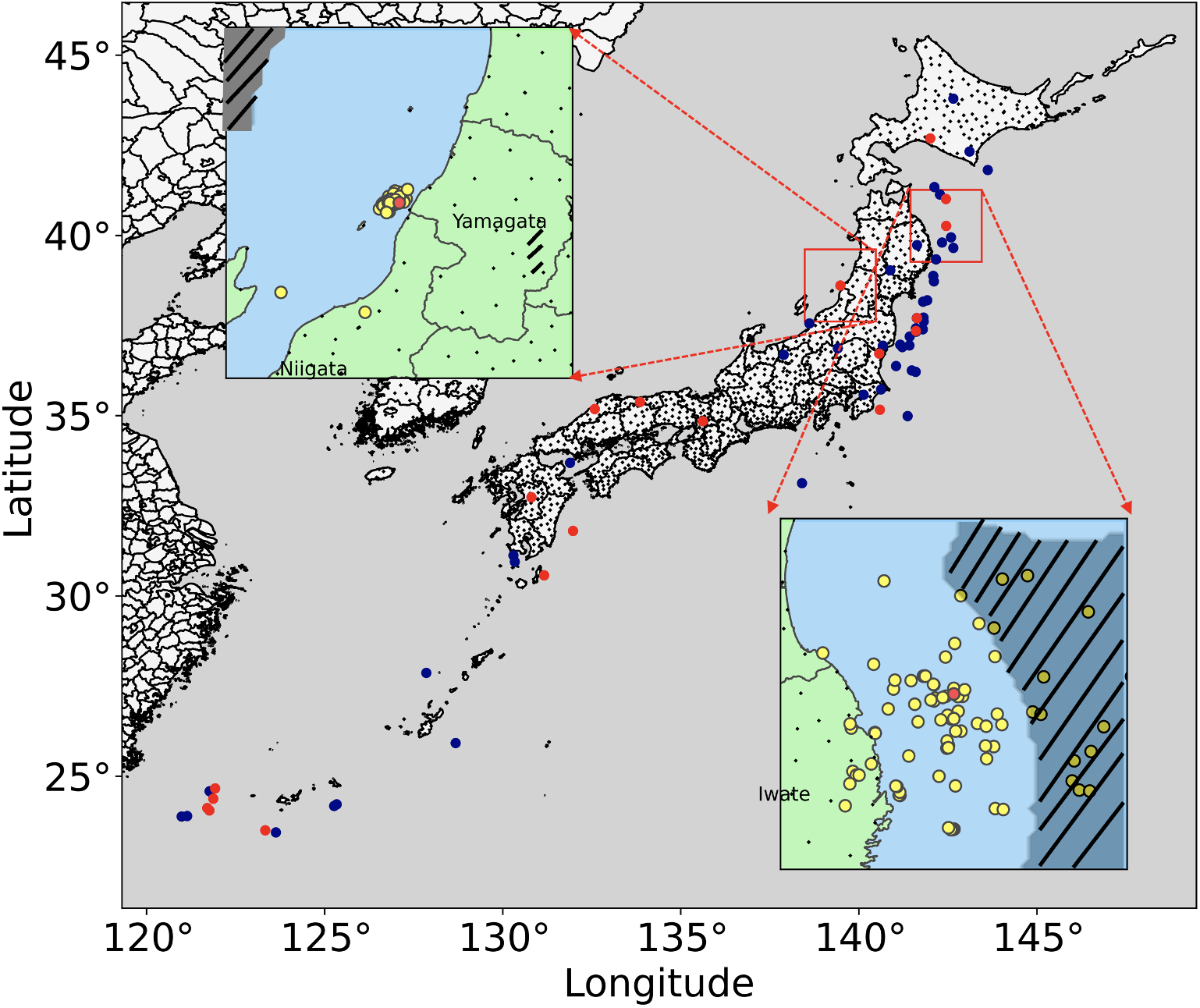}
    \caption{The location of the GPS stations (black dots) in Japan. The training set contains $48$ mainshocks  occurred before 2015/01/01 (blue circles). The test set contains $19$ mainshocks  occurred after 2015/01/01 (red circles) Insets: The $250$ km $\times \, 250$ km maps centered around two mainshocks:  the 2019 Tsuruoka earthquake (label 17)(upper left) and the 2017 offshore Hachinohe earthquake (label 7) (lower right).
    The aftershocks with magnitude larger than $4$ within $45$ days are depicted in yellow. The hatched regions correspond to masked cells, far from  GPS stations.}
    \label{fig:japan_map}
\end{figure}

The earthquakes used in this study lie in the longitude interval $[123 ^\circ,148 ^\circ]$ and latitude interval $[22 ^\circ,46 ^\circ]$ and are extracted from the JMA catalogue. 
We consider as a mainshock an earthquake with magnitude larger than $6$ and associate to each of them the list of the aftershocks that occurred within the next $45$ days, in the surrounding region (square box of $250$ km of width). Only big aftershocks with magnitude larger than $4$ are retained.  
Note that, in the case of two earthquakes with magnitude larger than $6$ lying within the same spatio-temporal window, only the first is labeled as a mainshock while the second is labeled as an aftershock of the former. 
The choice of $45$ days comes as a compromise between missing aftershocks (if the window was chosen shorter) or discarding sequences (if the window was chosen longer).
In Fig.\ref{fig:japan_map} we indicate the position of all mainshocks: in blue are earthquakes that occurred before 1st January 2015 and used for training the neural network, in red are the earthquakes that occurred after 1st January 2015 and build the test set to assess the quality of our results.

GPS data are extracted from the Nevada Geodetic Laboratory repository (see \citeA{Blewitt2018}). We use 24-hour (daily) final solutions in the frame 'IGS14', format 'tenv3'. Each GPS station, provides its daily position $\vec{r}(t) = (e(t), n(t))$ (here $e(t)$ stays for the longitude (eastward) and $n(t)$ for  latitude (northward) at the day $t$). Altitude measurements are also present, but are more noisy and thus discarded from our study. Hence, the daily displacement of the station  in the east-north plane is simply obtained as $\vec{v}(t)=\vec{r}(t)-\vec{r}(t-1)$.

\begin{figure}[H]
    \centering
    \includegraphics[height=8cm]{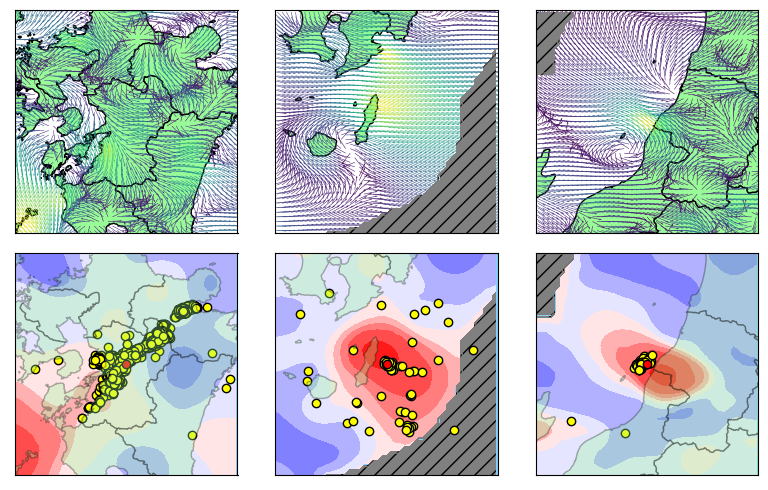}
    \caption{Top: From left to right, surface displacement map of the mainshocks $1,15,17$ (see Tab.(\ref{listtest})).  This is the input of the neural network. Down: the final prediction $\pi_i(x,y)$.  This is the output of the neural network, obtained with the MAX protocol.}
    \label{fig:maps}
\end{figure}

The input of the neural network are surface displacement maps of $250$ km $\times \, 250$ km centered at the mainshock and discretized in cells of $5$ km $\times \,5$ km. 
This choice of $5$ km is identical to that of \cite{DeVries2018} and allows to avoid interpolation problems that arise when choosing overly fine-grained discretization. We take a box of $250$ km to avoid missing out too many aftershocks, while avoiding an absurdly large window where most cells would be inactive.
The maps are obtained from the interpolation of the GPS measurements available only at the station positions. How to perform this interpolation is a classical issue in geophysics where the study of long time series of GPS data is important to estimate the strain rates induced by relative fault motion.  Classical interpolation schemes are based on least squares methods (weighted linear regression or biharmonic spline, see \cite{shen2015optimal}). Here we use instead a different approach, recommended in \cite{sandwell2016interpolation} and based on the solution of the 2D  Navier-Cauchy equations at the  equilibrium, describing the displacement of a (2D) solid at rest. In Japan the network of GPS is dense, but many earthquakes occur in the small islands or close to the sea where GPS measurements are only partially available. 
Because an interpolation cannot extrapolate displacements far away from stations, we do not include remote regions in our study. Concretely, cells for which the closest GPS station is further away than $d^*=86 \rm{km}$ are masked, i.e. they are excluded both from learning as well as forecasting.  This choice is a compromise between quality of data (better quality with smaller $d^*$) and quantity of data (larger amount with larger $d^*$).
In Fig.\ref{fig:maps} top  show some surface displacements maps at the  day of the mainshock,  the  input of the  neural network. For a given  cell $(x,y)$, the displacement $\vec{v}(x,y)=(v_e(x,y), v_n(x,y))$ is represented by an arrow (for the direction) and a color (for the modulus). Each cell has an associated binary label $A(x,y)$ which is $1$ if at least one aftershock with magnitude $m>4$ occurs in the $45$ days following the mainshock. 

We disregard sequences in remote regions, far from GPS stations\footnote{We keep the surface displacement maps with at least 3 stations employed for interpolation and with at least one inside the $250 \rm{km} \times 250 \rm{km}$ window.}, and select $67$ seismic sequences  divided in $N_{\rm tr} = 48$ sequences for the training set and $19$ for the test set.  
For each sequence, the goal is to predict the binary labels $A(x,y)$ given the surface displacement maps  $\vec{v}(x,y)$ as input.
Because the displacement induced by the earthquake is sometimes recorded on the next day, for each mainshock we actually provide two surface displacement maps: the one at the day of the mainshock as well as the one of the following day.
Thus, aftershocks occurring the immediate day after the mainshock are discarded from the prediction as well.
This prediction problem is clearly cast as a standard machine learning task. We deal with an input image of $50 \times 50$ cells of 4 channels each (the two components of the GPS displacement and the two days), and attempt to perform image segmentation, or in other words, binary classification of each cell. We use a Convolutional Neural Network (CNN), which is a neural network particularly suited for Computer Vision tasks, as it is translationally equivariant by construction \cite{726791,NIPS2012_c399862d}: this means that a shift in the input image induce an identical shift in the output.  The loss function is the Binary Cross Entropy loss, a classical choice for binary classification.
The input vectors of the displacement map are encoded using polar coordinates rather than cartesian ones, and the first layers of the CNN deal with the radial and angular part of the signal separately. The last layers combine these inputs to produce a scalar output, i.e.~a probability map $\pi(x,y) \in[0,1]$ of aftershock occurrence.
To face the issue of having such a small training set, we use two techniques. 
First we perform data augmentation, a classical technique: to each displacement map we apply all the plane transformations that leave the map a square, namely all possible distinct combinations of $4$ rotations of $0^\circ, 90^\circ, 180^\circ,270^\circ, $ and axial flip (8 possible variations in total). 
Second and most importantly, instead of using a single neural network trained with all the maps we construct an ensemble of neural networks. This ensemble procedure is used to have a more robust prediction, as the training set is small \cite{dietterich2000ensemble}. In practice, all the networks have identical architecture, but they are trained with $N_{\rm tr}-1$ independent mainshocks (hence two networks of the ensemble differ by a single mainshock). As a result, for each mainshock $i$ on the test set, we obtain an ensemble of $N_{\rm tr}$ predictions, indexed by $k=1...N_{\rm tr}$ :
\begin{align}
\{ \pi_{i}^{(1)}(x,y),\dots,\pi_{i}^{(k)}(x,y),\dots, \pi_{i}^{(N_{\rm tr})}(x,y)  \}
\end{align}
 To determine the final prediction $\pi_i(x,y)$ we use three protocols: 
 \begin{itemize}
\item The most cautious protocol is  MAX, where,  for each cell $(x,y)$, we select the maximal probability of the ensemble,  $\pi_i(x,y)=\max_k \left[\pi_{i}^{(k)}(x,y) \right]$. 
\item The  less cautious protocol is  MIN,  where we select the minimal probability of the ensemble,  $\pi_i(x,y) =\min_k\left[ \pi_{i}^{(k)}(x,y)\right]$. 
\item Finally, at the intermediate level, we define  MEDIAN where $\pi_i(x,y)$ is the median value among the $N_{\rm tr}$ sorted values of the ensemble.
\end{itemize}
We discuss in the next section how these different protocols perform  on the test set.
In the Appendices we provide the details of the interpolation method based on the 2D Navier-Cauchy equations and on the network architecture and training procedure. Moreover we provide the catalogue of the earthquakes employed in this study and all the Python codes for downloading and manipulating GPS data as well as training and testing the neural network. 
\section{Results}

\begin{figure}[H]  
    \centering
    \includegraphics[height=8cm]{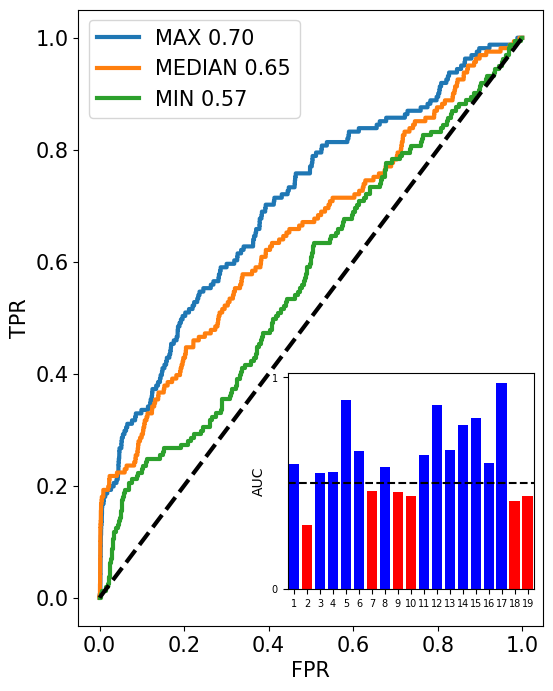}
    \includegraphics[height=8cm]{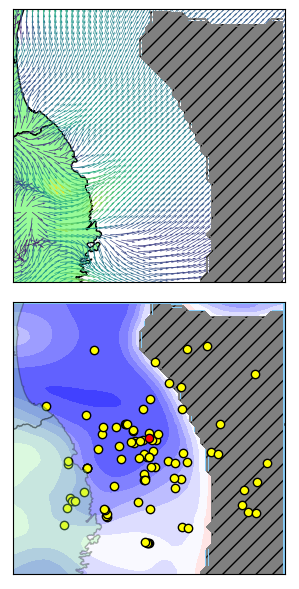}
    \caption{Left: the ROC curves obtained with the three protocols MAX, MEDIAN, MIN, averaged over the $19$ mainshocks of the test set. Inset: the AUC scores computed for each sample of the test set (MAX protocol). Right top: Surface displacement map of the mainshock $7$ (see Tab.(\ref{listtest})). Right down:  final prediction $\pi(x,y)$  (MAX protocol). In this case the network fails to produce a satisfying prediction. Even though 13 GPS stations were used to construct the surface displacement map, most of the events happen in the sea, where the signal quality is poor. }
    \label{fig:test_auc}
\end{figure}

The goal of this study is to use the GPS surface displacement map of the mainshock $i$ to predict if, within $45$ days, at least  an aftershock has occurred inside each cell $(x,y)$ of area $25 \rm{km}^2$. To  convert the continuous probability  $\pi_i(x, y)$ in a deterministic event with probability $0$ or $1$ we introduce a global threshold $\pi_{\rm th}$. Hence, we call positive a cell with $\pi_i(x,y) > \pi_{\rm th}$.
We divided the positive cells in two categories: 
\begin{itemize}
\item[(i)] the true positive are the positive cells where the prediction is correct. It is convenient to normalize their number by the total number of cells where at least an aftershock has occurred and obtain the true positive rate (TPR).
\item[(ii)] Similarly, the false positive are the positive cells  where the prediction is wrong. Their number should be normalized with  the total number of cells without aftershocks to obtain the false positive rate (FPR).
\end{itemize}
To assess the quality of our classifier we use the {\em receiver operating characteristic} (ROC) curve \cite{forecastVer,zechar2010evaluating}, obtained by  varying the parameter $\pi_{\rm th}$. The score associated to a ROC curve is usually given by the area under the curve (AUC). For example, the ROC curve of a random classifier is the line $\rm{FPR}=\rm{TPR}$ with an AUC of $0.5$; an AUC above $0.75$ is considered a very good classifier. In Fig. \ref{fig:test_auc} left the ROC curves obtained from the test set are shown. All protocols outperform the random classifier. The best protocol is the MAX, with an AUC of $0.7$. A fingerprint of the robustness of our results is that the MEDIAN protocol has a very similar ROC curve, proving  the majority of the networks in the ensemble tends to agree. Another fingerprint of such robustness is that the output probabilities $\pi_i(x,y)$ after each mainshock  are broadly distributed with cells with a very low probabilities $\sim 0.2$ and cells with high probabilities $\sim 0.8$. This means that the GPS displacement maps have a great forecasting potential because in the case of non-informative input data the output probabilities will have values narrow distributed around $\sim 0.5$.

To get further understanding in  the  inset of Fig. \ref{fig:test_auc} left  the AUC obtained for each mainshock of the test set are shown. Most of  them have a very good score, but a fraction of them perform worst than a random classifier. A representative mainshock belonging to this class is the 2017 Hachinohe (label $7$)  shown in Fig. \ref{fig:test_auc} right. In this case the mainshock is offshore quite close to the masked cells and thus in a region which  is quite far from the GPS stations. By inspecting  Tab.\ref{listtest} we observe  that a bad performance occurs either in  maps where the number of GPS stations used for the interpolation is small (labels $2$, $9$, $10$, $19$) either in situations where is located close to the masked cells (labels $7$ and $18$).

\section{Conclusion}
In conclusion our results show that this machine learning method based solely on GPS data is very promising. 
The principal limitation of the performance concerns the density of the GPS station which is high in Japan but much smaller in other seismic regions such as California, Italy, Greece, etc. \\
Our study is based only on the displacement map of the day of the mainshock. However, recent studies \cite{perfettini2004postseismic,perf2007,perf2005} have shown that large earthquakes are followed by an increased afterslip in the regions where the aftershock rate is very high as also confirmed by numerical models \cite{petrillo2020influence,lippiello2019fault}. It will be interesting to explore if the GPS data available after the mainshock can be used to improve the aftershock forecasting.
\\
Furthermore, for concrete improvement of the current aftershock prediction methods (ETAS models), one should consider a combination of past catalog data (ETAS approach) and recent surface deformation (GPS data processed with a CNN).

\acknowledgments
 V.M.S. thanks the valuable feedback from the conference "1ères Journées du GDR IDE "Interaction Désordre Élasticité" - 2022" where this work in preparation was presented. 
 G.P. acknowledges support from MEXT
Project for Seismology TowArd Research innovation with Data of Earthquake (STAR-E Project), Grant Number: JPJ010217. 
The GPS data used in the study is automatically downloaded from the repositories of the Nevada Geodetic Laboratory \cite{Blewitt2018} (http://geodesy.unr.edu/). We thank each independent work that processed the data from the individual GPS stations employed.
The earthquake catalog used in this study is produced by the Japan Meteorological Agency, in cooperation with the Ministry of Education, Culture, Sports, Science and Technology. The catalog is based on seismic data provided by the National Research Institute for Earth Science and Disaster Resilience, the Japan Meteorological Agency, Hokkaido University, Hirosaki University, Tohoku University, the University of Tokyo, Nagoya University, Kyoto University, Kochi University, Kyushu University, Kagoshima University, the National Institute of Advanced Industrial Science and Technology, the Geographical Survey Institute, Tokyo Metropolis, Shizuoka Prefecture, Hot Springs Research Institute of Kanagawa Prefecture, Yokohama City, and Japan Agency for Marine-Earth Science and Technology.

\begin{appendix}
\section{Computation of the surface displacement map via  the 2D Navier-Cauchy interpolation}
 GPS measurements are available only at the station positions $\{x_s, y_s \}_{s=1}^N$ ($x_s$ is the longitude, $y_s$ is the latitude). Hence, at the day of the mainshock we dispose  for each station of its displacement vector $\vec{v}(x_s,y_s)=(v_e(x_s,y_s), v_n(x_s,y_s))$. To construct the map we  need to compute the displacement vector of each cell $(x,y)$.
Following \cite{sandwell2016interpolation} we model the surface of the earth as $2D$ thin elastic sheet. The displacement satisfies the 2D Navier-Cauchy equations:
\begin{eqnarray}     
   \frac{2}{1-\nu} \partial_x^2 v_e + \frac{2\nu}{1-\nu} \partial_{xy}^2 v_n + \partial_y^2 v_e + \partial_{xy}^2 v_n & = -f_e \delta(x-x_0) \delta(y-y_0) &  \nonumber \\
   \partial_{xy}^2 v_e + \partial_{x}^2v_n + \frac{2\nu}{1-\nu} \partial_{xy}^2 v_e + \frac{2}{1-\nu} \partial_y^2 v_n & =-f_n \delta(x-x_0) \delta(y-y_0)  &
\end{eqnarray}

Here $\nu$ is the Poisson ratio which we set $1/2$ assuming the medium isotropic, $f_e$ and $f_n$ are the point forces acting on the sheet at a location $(x_0,y_0)$ (the forces are rescaled in the unit of the bulk modulus which we assume constant in space). The solution of these linear equations can be written as (see \cite{sandwell2016interpolation}):
\begin{eqnarray}
\label{eqn:nc2d_direct}
    v_e(x, y) & = & q(x-x_0, y-y_0) f_{e}+w(x-x_0, y-y_0) f_{n} \nonumber \\
    v_n(x, y) & = & w(x-x_0, y-y_0) f_{e}+p(x-x_0, y-y_0) f_{n}
\end{eqnarray}
where $q, p$ and $w$ are the Green's functions:
\begin{eqnarray}
    q(x,y) & = &(3-\nu) \ln(r) + (1+\nu) \frac{y^2}{r^2} \\
    p(x,y) & = &(3-\nu) \ln(r) + (1+\nu) \frac{x^2}{r^2} \\
    w(x,y) & = & -(1+\nu) \frac{xy}{r^2} 
\end{eqnarray}
with $r=\sqrt{x^2+y^2}$. To avoid the logarithmic divergence  the Green's functions are regularized when $r=0$ by replacing $\ln(r) \to \ln(r+a)$ with the constant $a=10$ km.

The interpolation strategy proposed in \cite{sandwell2016interpolation} is composed by two steps:
\begin{itemize}
\item First we assume that the point forces acting on the sheet are located at the station positions. To determine their value, we invert the following  $2 N\times 2N $ linear system
\begin{equation}
\label{eqn:nc2d_direct_measured}
  \begin{bmatrix}
v_e(x_{s}, y_{s}) \\
v_n(x_{s}, y_{s}) 
\end{bmatrix} = \underbrace{ \begin{bmatrix}
& q(x_s-x_{s'}, y_s-y_{s'}) & w(x_s-x_{s'}, y_s-y_{s'}) \\
& w(x_s-x_{s'}, y_s-y_{s'})  & p(x_s-x_{s'}, y_s-y_{s'})  
\end{bmatrix}}_G
\begin{bmatrix}
f_e(x_{s'}, y_{s'}) \\
f_n(x_{s'}, y_{s'}) 
\end{bmatrix}
\end{equation}
To invert the linear system matrix we use SVD decomposition $G=U\Sigma V^*$ and keep the $80\%$ of the eigenvalues of the matrix $\Sigma$. This allows to a smooth the final map and avoid singularities in the inversion. 
\item Second, we use equation  (\ref{eqn:nc2d_direct})  with the forces computed in the first step to find the displacement vector $(v_e(x, y), v_n(x, y))$ at the cell  positions $(x,y)$. 
\end{itemize}

\section{Network architecture and training procedure}




The building block of Convolutional Neural Networks (CNNs) is indeed the convolution operation. Given an image $I(x,y)$, the convolution with a \textit{filter} of size $2k+1 \times 2k+1$, denoted by $W_{ij}$, produces a new 'image'  $(W \star I)$:
\begin{equation}
    (W \star I)(x,y) = \sum_{i,j=-k}^k W_{i,j} I(x+i, y+j)
\end{equation}
In practice, we use mostly $k=2$, i.e. a filter of $5\times 5$. 
The parameters $W_{i,j}$ are learned during training.
As the convolution operation is linear, one must add a non-linearity after the convolution operation in order to make the network \textit{expressive}. A popular choice is a ReLU (rectified linear unit) function, which is nothing but $\textrm{ReLU}(z)=\max(z,0)$. In practice the ReLU is applied, pixel by pixel, on the output of the convolution.
A convolution followed by a non linearity, $\textrm{ReLU}(W \star I)$, defines a Convolution Block. A CNN consists in a series of Convolution Blocks stacked one after the other.

\begin{figure}[H]
    \centering
    \includegraphics[height=6cm]{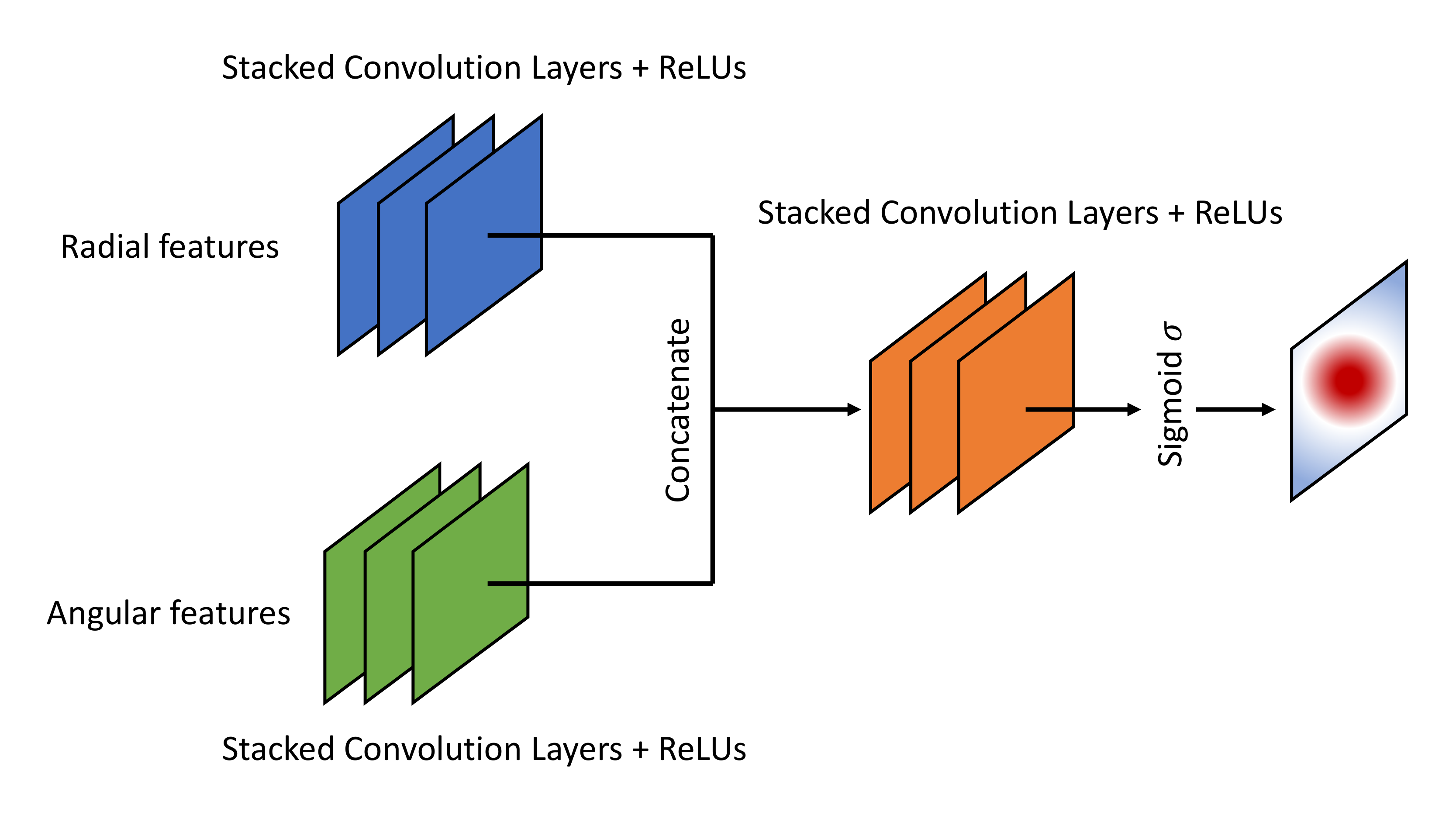}
    \caption{Scheme for the network architecture. The two convolutional blocks on the left are concatenated and passed to a last convolutional block that produces the output probabilities.}
    \label{fig:nn_schema}
\end{figure}

As mentioned in the main text, we use the polar representation of surface displacement maps as input, obtained as following:
\begin{eqnarray}
    v_r(x,y) & = & \sqrt{v_e(x,y)^2+v_n(x,y)^2} \nonumber \\
    s(x,y) & = & v_n(x,y)/v_r(x,y) \nonumber \\
    c(x,y) & = & v_e(x,y)/v_r(x,y) \nonumber
\end{eqnarray}
$v_r(x,y)$ is thus the magnitude of the displacement and $s(x,y)$ and $c(x,y)$ are the sine and cosines of the displacement vector on the $2D$ plane. To process the inputs, we use two CNNs made of 3 convolutional blocks each. The radial input is fed to one, the angular to the other. The outputs from the two networks are put together and fed to a final CNN with 3 blocks. The last block in this case does not use a ReLU function but a sigmoid $\sigma(z)=1/(1+e^{-z})$ that has domain on $[0,1]$ (i.e. represents a probability). Given all these details, we can describe using three equations our network:
\begin{eqnarray}
    o_r(x,y) & = & \textrm{ReLU}(W_{3,r} \star \textrm{ReLU}(W_{2,r} \star \textrm{ReLU} (W_{1,r} \star v_r(x,y) ))  ) \\
    o_a(x,y) & = & \textrm{ReLU}(W_{3,a} \star \textrm{ReLU}(W_{2,a} \star \textrm{ReLU} (W_{1,a} \star \{ s(x,y), c(x,y) \} ))  ) \\
    \pi(x,y) & = &  \sigma(W_{3} \star \textrm{ReLU}(W_{2} \star \textrm{ReLU} (W_{1} \star \{o_r(x,y), o_a(x,y)\} ))  )
\end{eqnarray}
The training of a neural network is nothing but the minimization of a Loss function by means of gradient descent, or some refined version of this classic optimization method, where the variables to be optimized are the weights of the network (the coefficients of the filters $W$).
The appropriate loss function for a binary classification is the binary cross entropy loss:
\begin{align}
  \mathcal{L} =  -\sum_{x,y,i}A_i(x,y) \ln( \pi_i(x,y))  + (1-A_i(x,y)) \ln (1-\pi_i(x,y))
\end{align}
where $i$ is the sample index (a sequence is a sample), $A_i(x,y)$ is the binary label to predict, 1 if cell $(x,y)$ has an aftershock and 0 elsewhere, and $\pi_i(x,y)$ is the predicted output probabilities (i.e. the whole neural network output for the sample $i$).
We mask the cells that are too far from any station (further away than $d^*=86km$) by taking them out from the sum.
The whole architecture is summarized in Fig.~\ref{fig:nn_schema}.

We now provide details on how to search the best CNN architecture, i.e. the procedure that led us to chose our machine learning model.
A priori, numerous architectures (different combinations of layers) can be considered, and the goal is to pick whichever performs best. 
However, because of the scarcity of data points, it's not trivial to measure the notion of ``works best''.
A naive approach would simply split the training set in a train and validation part, using the classification score (as measured e.g.~by the AUC) and loss function values measured on the validation data as estimates of the performance of a given architecture. 
However, this is very much not robust: given the small number of train samples, and thus also of validation samples, this would result in an essentially random choice, that would crucially depend on which samples where picked for the validation set.
\\
Thus, for each architecture or set of hyper-parameters we want to try, we perform \textit{leave-one-out cross-validation}: 
we build $N_{tr}$ train/validation splits, for each one we train the model using $N_{tr}-1$ points and measure the validation score on the left-out point.
The average performance over all splits is our measure of validation score, and the spread of scores between splits is a measure of the robustness of the model: a large spread means the model is not very robust.
\begin{figure}
    \centering
    \includegraphics[height=6cm, trim={0 0 0 1.4cm}, clip]{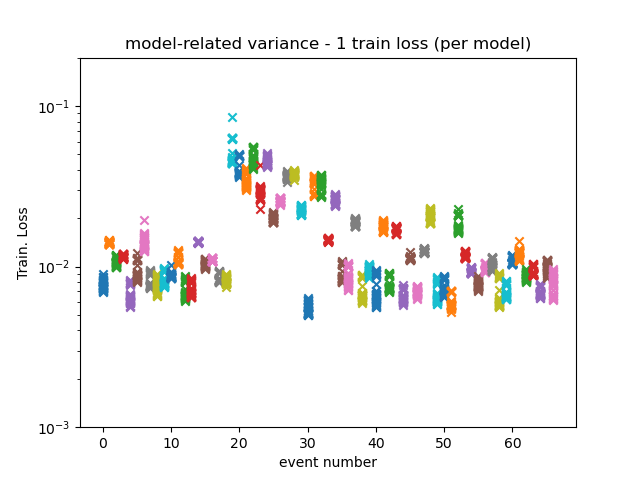}
    \caption{Validation scores of all splits, for each event in the train/validation set.  Each event corresponds to an abscissa and is assigned a random color for clarity. 
    We note that for a given event (fixed abscissa), the spread of scores of different models is small, with respect to the spread between events.}
    \label{fig:val_bars}
\end{figure}
In Fig.~\ref{fig:val_bars} we show that most of the randomness between splits comes from that data points being different, not from the models.
In our search of optimal model, we tried to get both large validation score and small spread between scores.

To reproduce the results in this work we refer to the code and data stored on GitHub \url{https://github.com/vicioms/gps_aftershocks_ml}. Moreover in a README file we provide the technical details of the training procedure. 
\section{Catalogue of the test set}
In the GitHub provided \url{https://github.com/vicioms/gps_aftershocks_ml}  the full list of mainshocks and aftershocks employed is available. For simplicity we list in Table \ref{listtest} the $19$  mainshocks of the test test. The blue color corresponds to mainshocks for which the forecasting works well, the color red is used for the others (see also the inset of Fig.2 left) . 
\begin{table}[H]
\caption{List of the $19$ earthquakes employed for the test.}
\vspace{0.3cm}
\centering
\begin{tabular}{cccccc}
Index & Code Name & Data & Mw & Epicenter & \#GPS Stations  \\ 
 \hline
  \hline
\textcolor{blue}{1} & \textcolor{blue}{Kumamoto} & \textcolor{blue}{2016/04/14} & \textcolor{blue}{6.5} & \textcolor{blue}{32.74\textit{N} 130.81\textit{E}} & \textcolor{blue}{66} \\
 \hline
\textcolor{red}{2} & \textcolor{red}{Yilan} & \textcolor{red}{2016/05/12} & \textcolor{red}{6.5} & \textcolor{red}{24.66\textit{N} 121.92\textit{E}} & \textcolor{red}{1} \\
 \hline
 \textcolor{blue}{3} & \textcolor{blue}{Yonakuni} & \textcolor{blue}{2016/06/24} & \textcolor{blue}{6.2} & \textcolor{blue}{23.50\textit{N} 123.33\textit{E}} & \textcolor{blue}{1} \\
 \hline
 \textcolor{blue}{4} & \textcolor{blue}{Kurayoshi} & \textcolor{blue}{2016/10/21} & \textcolor{blue}{6.6} & \textcolor{blue}{35.38\textit{N} 133.86\textit{E}} & \textcolor{blue}{41} \\
 \hline
 \textcolor{blue}{5} & \textcolor{blue}{Namie} & \textcolor{blue}{2016/11/22} & \textcolor{blue}{7.4} & \textcolor{blue}{37.35\textit{N} 141.60\textit{E}} & \textcolor{blue}{23}\\
 \hline
 \textcolor{blue}{6} & \textcolor{blue}{Daigo} & \textcolor{blue}{2016/12/28} & \textcolor{blue}{6.3} & \textcolor{blue}{36.72\textit{N} 140.57\textit{E}} & \textcolor{blue}{46} \\
 \hline
 \textcolor{red}{7} & \textcolor{red}{Hachinohe} & \textcolor{red}{2017/09/27} & \textcolor{red}{6.1} & \textcolor{red}{40.27\textit{N} 142.45\textit{E}} & \textcolor{red}{13} \\
 \hline
 \textcolor{blue}{8} & \textcolor{blue}{Misawa} & \textcolor{blue}{2018/01/24}  & \textcolor{blue}{6.3} & \textcolor{blue}{41.01\textit{N} 142.45\textit{E}} & \textcolor{blue}{6} \\
 \hline
 \textcolor{red}{9} & \textcolor{red}{Hualien} & \textcolor{red}{2018/02/04} & \textcolor{red}{6.5} & \textcolor{red}{24.12\textit{N} 121.70\textit{E}} & \textcolor{red}{1} \\
 \hline
 \textcolor{red}{10} & \textcolor{red}{Hualien} & \textcolor{red}{2018/02/08} & \textcolor{red}{6.2} & \textcolor{red}{24.05\textit{N} 121.76\textit{E}} & \textcolor{red}{2} \\
 \hline
 \textcolor{blue}{11} & \textcolor{blue}{Matsue} & \textcolor{blue}{2018/04/09} & \textcolor{blue}{6.1} & \textcolor{blue}{35.18\textit{N} 132.59\textit{E}} & \textcolor{blue}{31}\\
 \hline
 \textcolor{blue}{12} & \textcolor{blue}{Osaka} & \textcolor{blue}{2018/06/18} & \textcolor{blue}{6.1} & \textcolor{blue}{34.84\textit{N} 135.62\textit{E}} & \textcolor{blue}{75} \\
 \hline
 \textcolor{blue}{13} & \textcolor{blue}{Chiba} & \textcolor{blue}{2018/07/07} & \textcolor{blue}{6.0} & \textcolor{blue}{35.16\textit{N} 140.59\textit{E}} & \textcolor{blue}{29} \\
 \hline
 \textcolor{blue}{14} & \textcolor{blue}{Chitose} & \textcolor{blue}{2018/09/06} & \textcolor{blue}{6.7} & \textcolor{blue}{42.69\textit{N} 142.01\textit{E}} & \textcolor{blue}{31} \\
 \hline
 \textcolor{blue}{15} & \textcolor{blue}{Nishinoomote} & \textcolor{blue}{2019/01/08} & \textcolor{blue}{6.0} & \textcolor{blue}{30.57\textit{N} 131.16\textit{E}} & \textcolor{blue}{15} \\
 \hline
 \textcolor{blue}{16} & \textcolor{blue}{Miyazaki} & \textcolor{blue}{2019/05/10} & \textcolor{blue}{6.3} & \textcolor{blue}{31.80\textit{N} 131.97\textit{E}} & \textcolor{blue}{20} \\
 \hline
 \textcolor{blue}{17} & \textcolor{blue}{Tsuruoka} & \textcolor{blue}{2019/06/18} & \textcolor{blue}{6.7} & \textcolor{blue}{38.61\textit{N} 139.48\textit{E}} & \textcolor{blue}{28} \\
 \hline
 \textcolor{red}{18} & \textcolor{red}{Namie} & \textcolor{red}{2019/08/04} & \textcolor{red}{6.4} & \textcolor{red}{37.71\textit{N} 141.63\textit{E}} & \textcolor{red}{25} \\
 \hline
 \textcolor{red}{19} & \textcolor{red}{Yilan} & \textcolor{red}{2019/08/08} & \textcolor{red}{6.4} & \textcolor{red}{24.37\textit{N} 121.87\textit{E}} & \textcolor{red}{3} \\
 \hline
\end{tabular}
\label{listtest}
\end{table}
\end{appendix}

\bibliography{bibliography}

\end{document}